\documentclass{elsart}
\usepackage{amssymb,graphicx}
\usepackage{amsmath}
\usepackage{psfrag,epsfig}

\begin{document}

\begin{frontmatter}

\title{Preparation and transport properties of non-hydrated Na$_{0.5}$CoO$_2$ single crystals}

\author[ZJU_P]{X. Z. Chen},
\author[ZJU_P]{Z. A. Xu\corauthref{cor1}}\ead{zhuan@css.zju.edu.cn},
\author[ZJU_P]{G. H. Cao},
\author[ZJU_P]{J. Q. Shen},
\author[ZJU_T]{L. M. Qiu},
and \author[ZJU_T]{Z. H. Gan}
\corauth[cor1] {Corresponding
author}
\address[ZJU_P]{Department of Physics, Zhejiang University,
Hangzhou 310027, P. R. China}
\address[ZJU_T]{Institute of Refrigeration and
Cryogenic Engineering, Zhejiang University, Hangzhou 310027, P. R.
China}

\date{\today}

\begin{abstract}
Single crystals of Na$_{0.5}$CoO$_2$ were obtained through a flux
method followed by de-intercalation of sodium. The
Na$_{0.5}$CoO$_2$ samples were found to be vulnerable to water in
the air and a hydration process in which H$_2$O molecules fill
oxygen vacancies in CoO$_2$ layers is suggested to be responsible
for the unusual vulnerability to water. The transport properties,
including resistivity (\emph{$\rho$}), thermopower (\emph{S}) and
Hall coefficient (\emph{R$_H$}), were studied in a temperature
range of 5-300 K. The compound shows a weak localization of
carriers just below 200 K and Co$^{3+}$-Co$^{4+}$ charge ordering
at about 30 K, a relatively lower temperature than previously
reported. The results seem to be quite different from those
previously reported for this system [Foo et al, Phys. Rev. Lett.
92 (2004) 247001]. Possible mechanism underlying this kind of
inconsistency is discussed.
\end{abstract}


\end{frontmatter}


Layered transition-metal oxide Na$_x$CoO$_2$ has attracted great
interest for its unusual behaviors due to strongly correlated
electrons in the past years. It has a crystal structure consisting
of layers of edge shared CoO$_6$ octahedra, between which Na$^+$
ions are inserted. Early researches have found an unusually large
thermoelectric power with Na content \emph{x}=2/3~\cite{ref1}.
Spin entropy carried by holes (Co$^{4+}$) hoping in a Co$^{3+}$
background has been regarded as the likely source of enhanced
thermopower~\cite{ref2}. Recently, superconductivity with a
transition temperature \emph{T$_c$} of about 5 K has been found in
water-intercalated compound
Na$_{0.35}$CoO$_2$-1.3H$_2$O~\cite{ref3}, which has been viewed as
a breakthrough in search of exotic superconductors other than the
copper oxides. Magnetic susceptibility studies~\cite{ref4,ref5}
have showed that Na$_{0.35}$CoO$_2$-1.3H$_2$O is an extreme type
II superconductor, just like high \emph{T$_c$} cuprates, and
extensive Co NMR/NQR studies~\cite{ref6,add1,add2} have shown
evidences for non-s-wave superconductivity. Besides the large
thermopower and superconductivity, very recent
research~\cite{ref7} found that as the sodium concentration
\emph{x} increases from 0.30 to 0.75, non-hydrated Na$_x$CoO$_2$
goes from a paramagnetic metal to a charge ordered insulator (at
\emph{x}=0.5), and then to a Curie-Weiss metal. The sudden
appearance of charge ordered state and unusual transport
properties of Na$_{0.5}$CoO$_2$ have made it a focus of attention.
The distinct properties of this compound have been determined to
be related to the Na$^+$ ion-vacancy ordering and consequential
Co$^{3+}$-Co$^{4+}$ charge ordering at low temperature.

In this paper, we prepared Na$_x$CoO$_2$ crystal samples with
\emph{x}=0.5 and studied the transport properties including
resistivity, thermopower and Hall coefficient. The samples were
found to be vulnerable to the water in the air and the reaction
between crystals and water seemed to bring significant influence
on transport properties of the samples. Furthermore, our results
on the temperature dependence of thermopower, magnetic
susceptibility and Hall coefficient were found quite different
from those in Ref.\cite{ref7}, which may throw further
complications on this compound.

The starting crystals of Na$_x$CoO$_2$ were grown by a flux method
following the procedure reported in Ref.~\cite{ref8}. To remove Na
content , the as-grown crystals were immersed in solutions
obtained by dissolving different amounts I$_2$ in acetonitrile.
After reacting for a month at ambient temperature, the samples
were washed several times with acetonitrile and then dried. The
crystal structure was characterized by X-Ray Diffraction (XRD).
The measurements of in-plane electrical resistivity \emph{$\rho$},
thermopower \emph{S}, Hall coefficient \emph{R$_H$} and magnetic
susceptibility \emph{M} were carried out using a Quantum-Design
PPMS-9 system. Unless mentioned otherwise, the transport
measurements were performed just after preparation procedure and
different measurements of a sample were taken as soon as possible
to ensure uniformity of the sample.

\begin{figure}
\includegraphics[width=12cm]{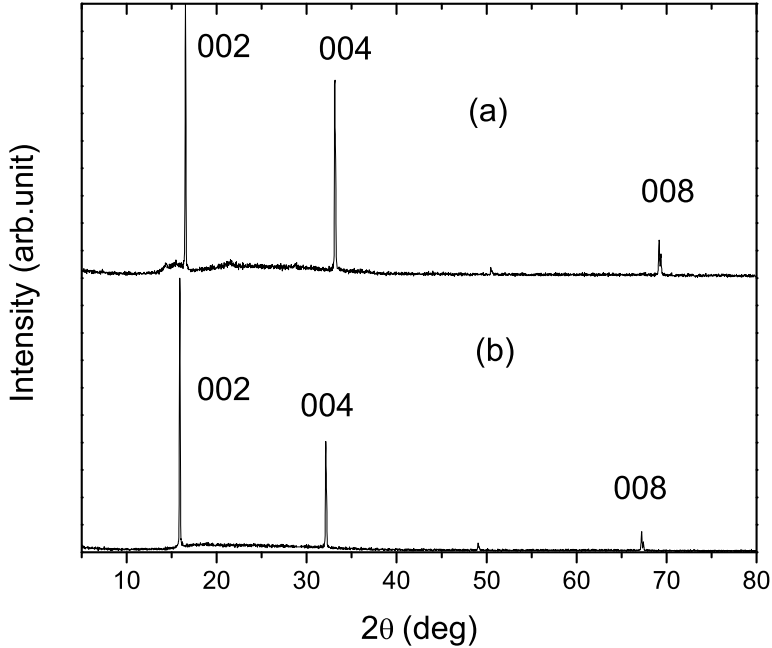} \caption{XRD patterns (Cu K¦Á radiation) for
Na$_x$CoO$_2$ crystals. (a) As-grown crystals Na$_{0.75}$CoO$_2$,
(b) de-intercalated crystal Na$_{0.5}$CoO$_2$.} \label{fig:fig1}
\end{figure}

Fig.~\ref{fig:fig1} shows the typical XRD patterns for the
as-grown crystal (a) and for de-intercalated crystal (b). From
these patterns, the crystals are considered to be of single phase.
The Na content \emph{x} of starting crystals was determined to be
0.75 from the c-axis parameter, using the calibration of the
relationship between c-axis parameter and Na content reported in
Ref ~\cite{ref7}. Na$_x$CoO$_2$ crystals with \emph{x} content
near 0.5 can be easily obtained by de-intercalation using
I$_2$-CH$_3$CN solution with excess I$_2$ concentrations for 30
days. The long reaction time assures the uniformity of Na content
of de-intercalated crystals. It should be mentioned that so long
as there is enough I$_2$ in the solution, the Na content \emph{x}
of de-intercalated crystals is always near 0.5 despite the
different I$_2$ concentrations, which means that formation of the
composition with \emph{x} near 0.5 is favored during the
de-intercalation of Na.

\begin{figure}
\includegraphics*[width=12cm]{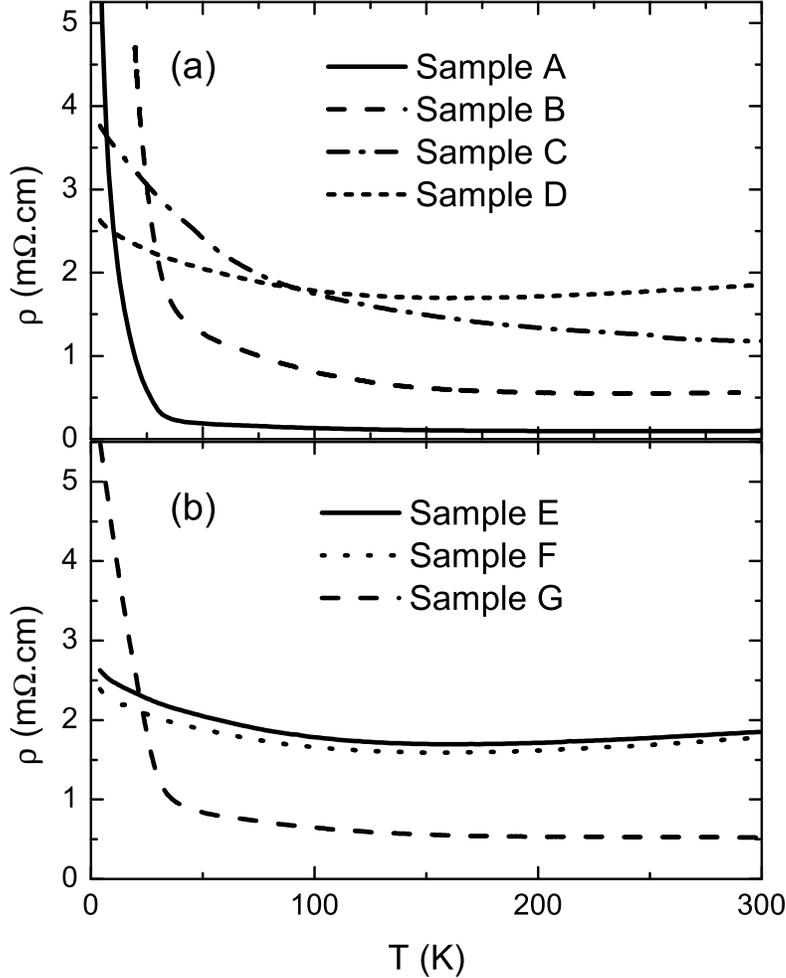} \caption{Temperature dependence of resistivity for
Na$_{0.5}$CoO$_2$ crystals with different water content. Sample A:
newly prepared samples(no water content), Sample B: expose Sample
A to dry air for 40 days, Sample C: expose Sample A to dry air for
80 days, Sample D: expose Sample A to dry air for 300 days, Sample
E: expose Sample A to humid air for 5 hours, Sample F: anneal
Sample E at 100 $\,^{\circ}\mathrm{C}$ for an hour, Sample G:
anneal Sample E at 250 $\,^{\circ}\mathrm{C}$ for an hour. All the
samples are of single phase, and no structural differences between
samples can be observed from XRD studies.} \label{fig:fig2}
\end{figure}

Na$_x$CoO$_2$ samples with low Na content (\emph{x} $<$ 0.35) are
easy to absorb water in the atmosphere and form hydrated
Na$_{x}$CoO$_2$-yH$_2$O with H$_{2}$O molecules being inserted
between CoO$_2$ layers. However, when \emph{x} comes as high as
0.45, the inserting of inter-layer H$_2$O is not
favored~\cite{ref9}. Meanwhile, our results show that
Na$_{0.5}$CoO$_2$ samples are vulnerable to the small content of
water in air. We have studied the time dependence of resistivity
of Na$_{0.5}$CoO$_2$ samples preserved in dry air as well as humid
air, as shown in Fig.~\ref{fig:fig2}. The original samples (Sample
A) were newly obtained by de-intercalation process mentioned
above, with c-axis lattice constant of 11.131 \AA, for which Na
content \emph{x} was estimated to be near 0.50. After each
measurement, the samples were preserved in dry air (cabinet with
humidity 35\%, Sample B, C, D) and humid air (Sample E). As can be
seen from Fig.~\ref{fig:fig2} (a), the fresh samples (Sample A)
show a transition below 50 K and the resistivity increases quickly
after this transition (The detailed behaviors of fresh samples
will be discussed below). The transition broaden and becomes less
evident and finally unobservable after the samples were kept in
air for a long period of time, which is shown in
Fig.~\ref{fig:fig2} (a) for Sample B, Sample C and Sample D. The
same variation of resistivity occurs much more rapidly when the
samples were put in humid air (Fig.~\ref{fig:fig2} (b), Sample E).
Sample F and G were obtained by annealing Sample E at
$100\,^{\circ}\mathrm{C}$ and $250\,^{\circ}\mathrm{C}$
respectively for an hour, and their resistivity curves were shown
in Fig.~\ref{fig:fig2}(b). It is very interesting that Sample F
(annealed at $100\,^{\circ}\mathrm{C}$) shows little difference in
resistivity with Sample E while Sample G has almost the same
behaviors in resistivity as Sample A, which means sintering at
$250\,^{\circ}\mathrm{C}$ has restore Sample E to its original
state (Sample A). It should be mentioned that, according to our
XRD studies, no observable structural changes corresponding to
this resistivity variation can be found and all the samples remain
single phase.

\begin{figure}
\includegraphics*[width=12cm]{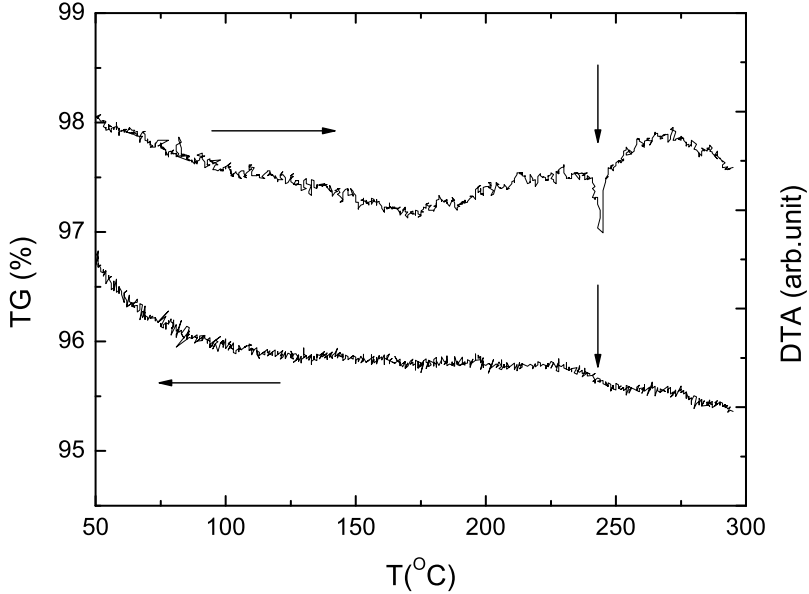} \caption{TG and DTA curves for water-incorperated
 Na$_{0.5}$CoO2 sample (Sample E). The arrows show the water loss at about 245
 $\,^{\circ}\mathrm{C}$.}
\label{fig:fig3}
\end{figure}

To give further insight into this process, we combined
thermogravimetric (TG) and differential-thermal-analysis (DTA)
studies with resistivity measurements. Fig.~\ref{fig:fig3} shows
TG and DTA curves for Samples E. From Fig.~\ref{fig:fig3}, two
weight losses can be observed, one below $100\,^{\circ}\mathrm{C}$
and one at about $245\,^{\circ}\mathrm{C}$, and the later is
accompanied by a negative peak in DTA curve. The first weight loss
below $100\,^{\circ}\mathrm{C}$ may come from the release of water
absorbed on the surface of samples and it makes no contribution to
the variation in resistivity. We believe that the weight loss at
$245\,^{\circ}\mathrm{C}$ is responsible for all resistivity
variations. According to some previous reports on superconducting
Na$_x$CoO$_{2}-y$H$_2$O, intercalation of water between CoO$_2$
layers will increasingly separate neighboring layers and thus
affect resistivity~\cite{SSC1}. This part of water will lose
quickly when samples were put in dry even at room temperature and
resistivity is restored~\cite{SSC1}. However, it's quite different
in our case. First, the small portion of water accounting for
resistivity variations in our samples will not be released from
the sample until $245\,^{\circ}\mathrm{C}$, a much higher
temperature than the usual temperature for releasing inter-layered
water. Second, only a very small proportion of water (about 0.015
H$_2$O per formula unit estimated from the weight loss at
$245\,^{\circ}\mathrm{C}$) can be absorbed into crystals even when
they are put in humid atmosphere for a long time. Third, no
structural changes were observed from XRD studies in spite of
dramatic changes in resistivity. All these differences make the
hydration process seem complicated. One of the probable mechanisms
is that H$_2$O molecules are scattered in CoO$_2$ layers instead
of being inserted between them~\cite{Cond1}. In such model, the
O$^{2-}$ ions of H$_2$O fill the oxygen vacancies in CoO$_2$ while
leaving H$^+$ suspended. If this is the truth, only a small
portion of water depending on the density of O$^{2-}$ vacancies
can be absorbed and definitely it will modify the periodic
potential, have influence on the Na$^+$ ion-Na vacancy ordering
and consequently contribute to transport properties of the system.
The density of O$^{2-}$ vacancies in CoO$_2$ layers can be created
during de-intercalation of Na content~\cite{CM1}, depending on the
preparation process. The special vulnerability of
Na$_{0.5}$CoO$_2$ samples to water makes consistent studies on
this system difficult, for instance, special care is required to
preserve samples and all measurements should be taken as soon as
possible to assure the uniformity of test samples.

\begin{figure}
\includegraphics*[width=12cm]{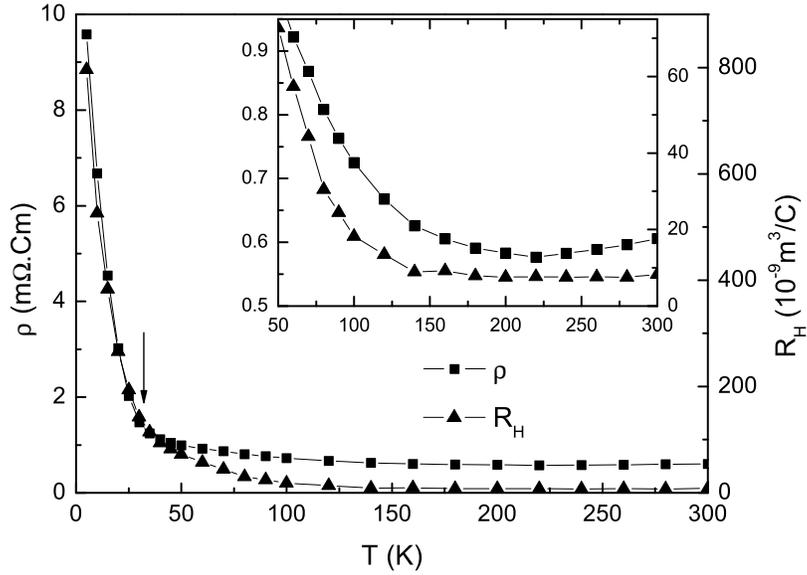} \caption{Temperature dependences of in-plane
resistivity \emph{$\rho$} and Hall coefficient \emph{R$_H$} for
Na$_{0.5}$CoO$_2$. The arrow show the transition at about 30 K.
Inset is the enlarged plot for $T \geq$ 50 K.} \label{fig:fig4}
\end{figure}

\begin{figure}
\includegraphics*[width=12cm]{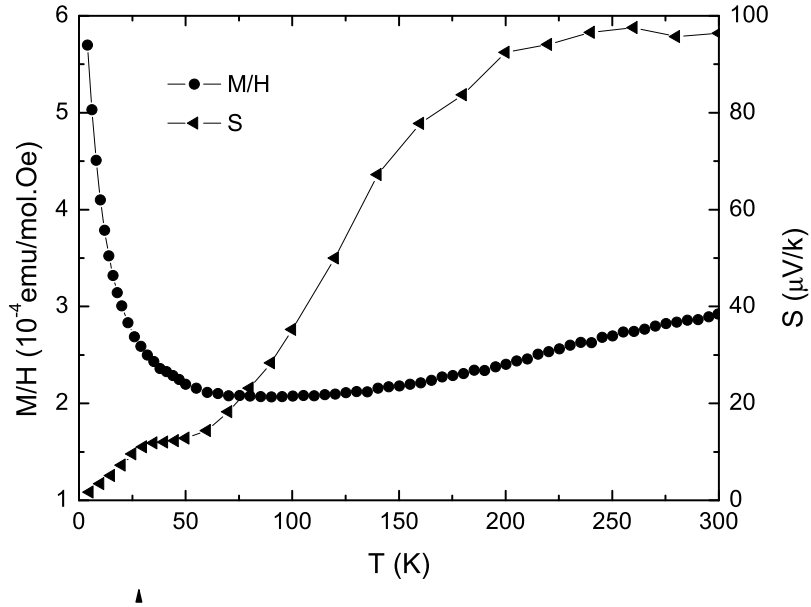} \caption{Temperature dependences of
thermopower \emph{S} and magnetic susceptibility \emph{M} (H
$\parallel$ c) for Na$_{0.5}$CoO$_2$. The arrow shows the
transition at 30 K .} \label{fig:fig5}
\end{figure}

 The transport measurements of Na$_{0.5}$CoO$_2$, including resistivity
\emph{$\rho$}, Hall coefficient \emph{R$_H$}, thermopower \emph{S}
and magnetic susceptibility \emph{M/H} were carried out in newly
prepared sample (Sample A) in a temperature range of 5-300 K.
Fig.~\ref{fig:fig4} shows the in-plane resistivity \emph{$\rho$}
and Hall coefficient \emph{R$_H$}, and Fig.~\ref{fig:fig5} shows
thermopower \emph{S} and magnetic susceptibility \emph{M/H}
(H$\parallel$c =1 T) as a function of temperature. \emph{$\rho$}
and \emph{R$_H$} show quite similar temperature dependence. They
both decreases gradually while temperature drops from room
temperature. However, below 200 K, they start to increase slowly
(see the enlargement of $\rho$-$T$ curve in the inset of
Fig.\ref{fig:fig4}). Correspondingly, \emph{S} starts to decrease
below 200K, as shown in Fig.\ref{fig:fig5}. A sharp transition in
transport properties can be observed at about 30 K. Below 30 K,
the increasing trend of \emph{$\rho$} and \emph{R$_H$} become much
rapid, associated with a quick increase in \emph{M} which is
similar to the result reported in Ref.~\cite{Chou}. \emph{S} drop
linearly to zero below 30 K (Fig.\ref{fig:fig5}). There is a tiny
cusp below 50 K in magnetic susceptibility and no anomaly can be
seen around 88 K.

According to previous study~\cite{ref7,Ord1}, superlattice of
ordered Na$^+$ ion-Na vacancy has been observed by electron
diffraction in Na$_{0.5}$CoO$_2$ even at room temperature and the
long-range ordering of Na$^+$ ion-vacancy at low temperature are
suggested for this compound by thermal conductivity measurement.
We presume that the changes starting around 200 K in
\emph{$\rho$}, \emph{R$_H$} and \emph{S} result from the ordering
of Na$^+$ ion-vacancy. The ordering of Na$^+$ ion-vacancy
introduces potential fluctuation which modulates electronic state
and consequently leads to Co$^{3+}$-Co$^{4+}$ charge ordering in
CoO$_2$ conduction layers. In our case, the Hall density at 100 K
for Na$_{0.5}$CoO$_2$ is 40 times smaller than that for
Na$_{0.75}$CoO$_2$(not shown in the paper). Below 30 K,
\emph{$\rho$}, \emph{R$_H$} and \emph{M} start to grow quickly,
these may come from Co$^{3+}$-Co$^{4+}$ charge ordering
transition.

However, there are many apparent differences between our results
and those reported in Ref.~\cite{ref7}. First,  no evident
transition in \emph{R$_H$} or \emph{S} at 87 K is observed from
our results. Second, the sharp transition in \emph{$\rho$} in our
case is around 30 K, while a sharp transition in \emph{$\rho$} at
53 K is observed in Ref.~\cite{ref7}. And third, the system in
Ref~\cite{ref7} shows evidence for electron-hole symmetry (Both
\emph{S} and \emph{R$_H$} fall toward zero as $T \rightarrow$ 0),
while in our case the system remains hole-like in the whole
temperature investigated and Hall density (1/e\emph{R$_H$})
instead of \emph{R$_H$} goes to zero as $T \rightarrow$ 0. It
should be noted that the behaviors of resistivity, thermopower and
Hall coefficient of our parent crystals ($x$=0.75) are very
similar to the those reported in Ref.~\cite{ref7} (not shown
here). Therefore the above differences for the samples with
$x$=0.5 seem to be surprising. Recalled that the Na$_{0.5}$CoO$_2$
crystals are vulnerable to water, we suggested that the oxygen
vacancies in the CoO$_2$ layers could account for these
differences. For the ideal composition of Na$_x$CoO$_2$ with
$x$=0.5, the ratio of Co$^{3+}$/Co$^{4+}$(eletron/hole) is 1. If a
certain amount of oxygen vacancies are introduced into the CoO$_2$
layers during preparing process(this is quite likely to happen in
Na$_{0.5}$CoO$_2$), it will destroy not only the electron-hole
symmetry, but also translation symmetry of the periodic potential.
Thus, we can expect that even a small number O$^{2-}$ vacancies
will have great influence on the transport properties of the
system. This is supported by the result that only 0.015 H$_2$O per
formula unit filling into O$^{2-}$ vacancies will substantially
change the resistivity of the system.

In summary, the single crystals of Na$_{0.5}$CoO$_2$ were prepared
and the transport properties were investigated. The unusual
vulnerability of the sample to water in the air was found and may
result from filling oxygen vacancies in CoO$_2$ layers with H$_2$O
molecules. The temperature dependence of resistivity, Hall
coefficient, and thermopower shows a weak localization of carriers
just below 200 K and Co$^{3+}$-Co$^{4+}$ charge ordering at about
30 K, a relatively lower temperature than previously reported. The
differences in transport properties between our results and those
reported in Ref.~\cite{ref7} may be caused by the oxygen vacancies
in CoO$_2$ layers. More studies are needed to discover the
underlying mechanism for the novel transport properties of this
system.

\textbf{Acknowledgements}
This work was supported by the National
Natural Science Foundation of China (Grant No. 10225417) and the
Ministry of Science and Technology of China (project:
NKBRSF-G1999064602).

\end{document}